\documentclass[sigconf]{acmart}
\AtBeginDocument{%
  \providecommand\BibTeX{{%
    \normalfont B\kern-0.5em{\scshape i\kern-0.25em b}\kern-0.8em\TeX}}}
    \usepackage{dblfloatfix} 
\usepackage{multirow}    
\usepackage{amsmath}
\usepackage{subcaption}
\usepackage{xcolor}
\usepackage[skins,most]{tcolorbox}
\usepackage{color, colortbl}
\usepackage[linesnumbered,ruled,vlined]{algorithm2e}

\usepackage{subcaption,siunitx,booktabs}

\newtcbtheorem{Answer}{\bfseries Answer}{enhanced,drop shadow={black!50!white},
  coltitle=black,
  top=0.1in,
  attach boxed title to top left=
  {xshift=1.5em,yshift=-\tcboxedtitleheight/2},
  boxed title style={size=small,colback=pink}
}{Answer}

\newtcolorbox[auto counter]{summary}[1][]{title={\bfseries Summary~\thetcbcounter},enhanced,drop shadow={black!50!white},
  coltitle=black,
  top=0.3in,
  attach boxed title to top left=
  {xshift=1.5em,yshift=-\tcboxedtitleheight/2},
  boxed title style={size=small,colback=pink},#1}

\newenvironment{RQ}{\vspace{2mm}\begin{tcolorbox}[enhanced,width=3.33in,size=fbox,
colback=blue!5,drop shadow southwest,sharp corners]}{\end{tcolorbox}}

\acmConference[Conference acronym 'XX]{Make sure to enter the correct
  conference title from your rights confirmation emai}{June 03--05,
  2018}{Woodstock, NY}

\makeatletter
\def\@copyrightspace{\relax}
\makeatother
\begin{document}

\title{Reducing the Cost of Training Security Classifier
(via Optimized Semi-Supervised Learning)}

\author{Rui Shu, Tianpei Xia, Huy Tu, Laurie Williams, Tim Menzies}
\affiliation{%
  \institution{North Carolina State University}
  \city{Raleigh}
  \state{North Carolina}
  \country{USA}}

\begin{abstract}
\textbf{Background}: Most of the existing machine learning models for security tasks, such as spam detection, malware detection, or network intrusion detection, are built on supervised machine learning algorithms. In such a paradigm, models need a large amount of labeled data to learn the useful relationships between selected features and target class. However, such labeled data can be scarce and expensive to acquire. \textbf{Goal}: To help security practitioners train useful security classification models when few labeled training data and many unlabeled training data are available. \textbf{Method}: We propose an adaptive framework called Dapper, which optimizes 1) semi-supervised learning algorithms to assign pseudo-labels to unlabeled data in a propagation paradigm and 2) the machine learning classifier (i.e., random forest). When the dataset class is highly imbalanced, Dapper then adaptively integrates and optimizes a data oversampling method called SMOTE. We use the novel Bayesian Optimization to search a large hyperparameter space of these tuning targets. \textbf{Result}: We evaluate Dapper with three security datasets, i.e., the Twitter spam dataset, the malware URLs dataset, and the CIC-IDS-2017 dataset. Experimental results indicate that we can use as low as 10\% of original labeled data but achieve close or even better classification performance than using 100\% labeled data in a supervised way. \textbf{Conclusion}: Based on those results, we would recommend using hyperparameter optimization with semi-supervised learning when dealing with shortages of labeled security data.
\end{abstract}

\maketitle

\section{Introduction}




When using machine learning to address security tasks, most existing models are built on supervised-learning algorithms. For example, many existing spam detection techniques~\cite{crawford2015survey,wu2018twitter}, malware detection techniques~\cite{souri2018state} or network intrusion detection systems~\cite{resende2018survey} train different classifiers to learn inherent relationships that exist between selected features and associated output class (i.e., label), namely abnormal and benign. Next, those classifiers are tested on unseen data for classification purposes. Thus, labeled data is necessary for training a helpful model in a supervised paradigm.

However, there are often cases when labeled security data is insufficient and expensive to collect, while a large set of unlabeled security data is available. To make good use of these unlabeled data, more practitioners resort to ways to annotate data to enlarge the size of labeled training data. Such process is referred as \textit{data annotation} or \textit{data labeling}. However, the work of data annotation is usually time-consuming and costly. For example, Tu et al.~\cite{tu2020better} in other domain (e.g., SE) task reported that manually reading and labeling 22,500+ GitHub commits requires 175 person-hours (approximately nine weeks), including cross-checking among labelers. Moreover, specific domain knowledge (e.g., security) is also required to ensure the high quality of annotated data.

\begin{table*}[!htbp]
\centering
\small
\caption{A list of prior work using semi-supervised learning in security tasks.}
\begin{tabular}{c|c|l}
\hline
\textbf{Publication} & \textbf{Year} & \multicolumn{1}{c}{\textbf{Brief Description}} \\ \hline \hline
\cite{le2021training} & 2021 & Use SSL to maximize the effectiveness of limit labeled training data for insider threat detection. \\ 
\cite{wang2015easeandroid} & 2015 & Model and automate the Android policy refinement process with SSL. \\ 
\cite{zhang2020label} & 2020 & Use label propagation to detect review spam groups. \\ 
\cite{nunes2016darknet} & 2016 & Help with classification task of identifying relevant products in darknet/deepnet marketplaces, etc. \\ 
\cite{wang2021collaboration} & 2021 & Provide a multi-label propagation based method for fraud detection. \\ 
\cite{alabdulmohsin2016content} & 2016 & Propose a method to estimate the maliciousness of a given file through a semi-supervised label propagation procedure. \\ 
\cite{taheri2020defending} & 2020 & Introduce a DL-based semi-supervised approach against label flipping attacks in the malware detection system. \\ 
\cite{pallaprolu2016label} &  2016 & Propose to use label propagation to discover infected Remote Access Trojans packets in large unlabeled data. \\ 
\cite{ni2015file} & 2015 & Use label propagation for malware detection. \\ 
\cite{kolosnjaji2016adaptive} & 2016 & Create a semi-supervised malware classification system that unifies views
of static and dynamic malware analysis. \\  \hline
\end{tabular}
\label{tbl:publications}
\end{table*}

Semi-supervised learning (SSL)~\cite{zhu2009introduction,zhu2005semi} can address the challenges as mentioned above from the algorithm perspective. SSL is a method that correlates the features in unlabeled data with labeled data and further generates pseudo-labels for the unlabelled data in an intuitive way. The newly labeled dataset (i.e., a mixture of both labeled and pseudo-labeled samples) is then used to train a model in a supervised manner. Without involving manual and expensive labor, SSL significantly reduces the cost of training models with a large labeled dataset instead of only a tiny portion of them. Table~\ref{tbl:publications} lists a sample list of prior work that uses semi-supervised learning in multiple security tasks. Among these work, there are two representative semi-supervised learning algorithms, i.e., \textit{label propagation}~\cite{zhu2002learning} and \textit{label spreading}~\cite{zhou2003learning}. Both algorithms use graph representation, compute data similarity between labeled and unlabeled data, and further propagate known labels through the edges of the graph to unlabeled data.

However, we observe that applications in Table~\ref{tbl:publications} rarely apply hyperparameter optimization on SSL algorithms. We argue that \textit{exploring suitable hyperparameter
configurations for semi-supervised learning algorithms would have an impact on the performance}. To validate this argument, we propose an adaptive framework called \textit{\textbf{Dapper}} that adopts hyperparameter optimization to control the configurations of the SSL algorithms. Beyond that, Dapper also explores the hyperparameter space of machine learning classifier (e.g., Random Forest). Furthermore, we also observe that, for some security datasets, the mixed training dataset (including labeled data and pseudo-labeled data) still suffers from class imbalance (which is a common issue in the security dataset). In this case, Dapper adaptively adds a tunable version of SMOTE~\cite{agrawal2018better} to rebalance the ratio between classes in the training datasets.



We evaluate Dapper with three real-world study cases, i.e. the Twitter spam dataset~\cite{chen20156}, malware URLs~\cite{mamun2016detecting} and CIC-IDS-2017 dataset~\cite{sharafaldin2018toward}. Our experimental results indicate that Dapper outperforms default SSL learners and optimized SSL learners. We can use as low as 10\% of the original labeled dataset but achieve close or even better classification performance (e.g., g-measure and recall) than using 100\% original labeled data. Based on those results, we recommend using Dapper framework when dealing with the shortage of available labeled data for security tasks. 

The remainder of this paper is organized as follows. We discuss background and related work in Section~\ref{sec:background} and our methodology in Section~\ref{sec:methodology}. We then report our experiment details in Section~\ref{sec:Evaluation}, including datasets, evaluation metrics, etc. Section~\ref{sec:results} presents our experiment results. We discuss the threats to validity in Section~\ref{sec:threats} and then we conclude in Section~\ref{sec:conclusion}.

\section{Background and Related Works}~\label{sec:background}

\subsection{Training Security Models Requires Labeled Data}

Diverse machine learning techniques have been widely applied in the cyber-security field to address wide-ranging problems such as spam detection, malware detection, and network intrusion detection. For example, in Twitter spam detection~\cite{chen20156}, machine learning algorithms use account-based features (e.g., the number of followers or friends) or message-based features (e.g., length of a tweet) to train useful models which are further used to predict other new spamming activities. In malware detection~\cite{mamun2016detecting}, several prior works analyze web URLs where malicious URLs are intended for malicious purposes such as stealing user privacy information. Security practitioners use machine learning techniques to classify malicious websites with features extracted from URLs such as URL tokens, length of URLs, etc. Another example is the network intrusion detection~\cite{mamun2016detecting} which endeavors to identify malicious behaviors in the network traffic. Machine learning-based techniques have gained enormous popularity in this field. They learn useful features from the network traffic and classify the normal and abnormal activities based on the learned patterns.

Security classification algorithms learn models from data, so insufficient training data can lead to low-quality models. But in many cases there are only a small number of labeled instances are available
(compared to a much large amount of unlabeled instances). For example, in a network intrusion detection scenario, network traffic of the monitored system is continuously generated with a large load, but the abnormal traffic, which is few and available only under malicious attacks~\cite{gharib2016evaluation}.

To make good use of unlabeled data, practitioners propose methods to annotate unlabeled data involving human efforts. However, the process of data annotation faces several challenges:

\begin{itemize}
    \item \textit{\textbf{Time-consuming and expensive}}. Due to the large volume of unlabeled data, much effort and time or finance is expected to be involved in the process, and it is not always an affordable solution. For example, manual labeling would cost \$320K and 39,000 hours to label GitHub issues of 50 projects as buggy or non-buggy~\cite{tu2020better}. 
    \item \textit{\textbf{Require domain knowledge}}. A lack of professional expertise is commonly the root cause of poor label quality. In the context of security, a person who lacks knowledge of security vulnerability, intrusion detection, malware, etc., hardly guarantees the right decision on tagging new data.
\end{itemize}

Moreover, existing data annotation methods can be mainly categorized into the following types:

\begin{itemize}
    \item \textbf{\textit{Manual}}. Manual data annotation works during the initial phase of a project when the data size is small and not complicated. This method is not salable for a large collection and might become overwhelmed and cause degraded label quality.
    \item \textbf{\textit{Crowdsourcing}}. Crowdsource data annotation is a better choice than manual labeling, which can be scalable with the help of platforms such as Amazon Mechanical Turk (MTurk)~\cite{mturk}. However, crowdsourcing can be expensive and can not guarantee the quality of the label.
    \item \textbf{\textit{Outsourcing}}. This method includes employing data labeling companies in low-cost markets. With extra QA processes and other solutions, label quality can be controlled and improved. However, outsourcing data annotation is still a manual process. 
    (albeit with a cheap cost of labor).
    \item \textbf{\textit{Interactive Learning}}. With methods such as active learning, rather than annotating all the data independently and simultaneously, only a fraction of the total number of data to be labeled~\cite{settles2008analysis}. With active learning, the expert users will label the most suitable data. 
\end{itemize}

In summary, a drawback with all the above methods is that they are all human-in-the-loop (HITL) methods. This kind of method raises the issue of requiring expertise of that human (and they might make mistakes). The rest of this paper explores fully automated techniques to avoid this issue.

\subsection{Semi-Supervised Learning}

Ground truth (i.e., labeled data) is often limited and costly to acquire when applying machine learning to security tasks. To address this problem, as shown in Table~\ref{tbl:publications}, some prior works in security have used {\em semi-supervised learning} to address such challenge. Specifically, semi-supervised learning (SSL)~\cite{zhu2009introduction,zhu2005semi} is a branch of machine learning algorithms that lies between supervised learning~\cite{hastie2009overview} and unsupervised learning~\cite{celebi2016unsupervised}. In supervised learning, the training dataset comprises only labeled data, which is to learn a function that can generalize well on the unseen data. Unsupervised learning only considers unlabeled data, where data points are grouped into clusters with similar properties. Semi-supervised learning combines both supervised learning and unsupervised learning, which uses a small amount of labeled data and a large amount of unlabeled data. The use of semi-supervised learning avoids searching labeled data or manually annotating unlabeled data. Major semi-supervised learning algorithms can be mainly categorized into the following two groups~\cite{silva2016survey}:

\textit{\textbf{Wrapper-Based Methods.}} Methods in this group use a supervised algorithm in an iterative way. During each iteration, a certain amount of unlabeled data is labeled by the decision function that is learned and incorporated into the training data. With the labeled data already available as well as its own prediction, the classification model is retrained for the next iteration. Two well-known representative methods in this group are \textit{self-training}~\cite{scudder1965probability} and \textit{co-training}~\cite{blum1998combining}. Self-training is a technique in which initially, a classifier is trained with the small amount of labeled data and then used to classify unlabeled data. The high confident unlabeled data as well as their predicted labels are added to the training dataset. The whole process is repeated either for a fixed number of iterations or until there are no high-confidence samples left in the unlabeled data. The co-training algorithm adopts an iterative learning process similar to self-training. It combines both labeled and unlabeled data under two-view setting. Initially, co-training trains two classifiers from each of the subset separately with limited labeled dataset. Then unlabeled data which the two classifiers have confident prediction will enlarge the labeled dataset for further training. This process repeats until a termination condition is met. The premise of co-training is that it assumes features can be split into two subsets and both subsets are conditionally independent given the class and sufficient to train classifiers by itself. Both method have several drawbacks. For example, in self-training, mistakes can re-enforce themselves. Co-training makes several assumptions, and only works well when conditional independence holds.

\textbf{\textit{Graph-Based Methods.}} Graph-based methods~\cite{van2020survey} create a graph that connects instances in the training dataset and propagates labels from labeled data to unlabeled data, through the edges of the graph. This process typically involves computing similarities between data instances. Consider the geometry of the dataset, which can be represented by an empirical graph $g = (V, E)$, where nodes $V = {1, ..., n}$ denote the training data and edges $E$ represents the similarities or affinity between adjacent nodes. Labels that assigned to the nodes in the graph can propagate along the edges of the graph to their connected nodes. The assumption behind the approach is that nodes with strong edges are more likely to share the same label. Two representative graph-based algorithms are \textit{label propagation}~\cite{zhu2002learning} and \textit{label spreading}~\cite{zhou2003learning}, which we will introduce in details in Section~\ref{sec:lpa} and Section~\ref{sec:lsa}. Compared with other semi-supervised learning algorithms, graph-based methods are fast and easy to use due to its linear time complexity, and therefore explored in this study.

\subsection{SMOTE}~\label{sec:smote}
Security data commonly suffer from data class imbalance issues. Many prior works propose methods to rebalance the ratio between classes to address this concern. SMOTE~\cite{chawla2002smote} (i.e., Synthetic Minority Oversampling Technique) is a widely used oversampling technique that works by randomly selecting samples from minority classes and choosing $k$ nearest neighbors for each chosen sample. A synthetic instance is created at a randomly selected point between each pair of chosen sample and its neighbor. The synthetic samples are added to the original dataset to balance the ratio between majority and minority classes. Agrawal et al. proposed an auto-tuning version of SMOTE, which is called SMOTUNED~\cite{agrawal2018better}. SMOTUNED adjusts several key parameters of SMOTE such as $k$ (the number of neighbors selected), $m$ (the number of synthetic samples to create), and $r$ (the power parameter for the Minkowski distance metric). SMOTUNED applies an evolutionary algorithm called differential evolution~\cite{storn1997differential} as the optimizer to explore SMOTE's parameter space. Our study uses SMOTUNED as our oversampling technique, but with a more novel optimizer introduced in the following subsection.

\begin{figure*}[!b]
\centering
\includegraphics[width=16.5cm]{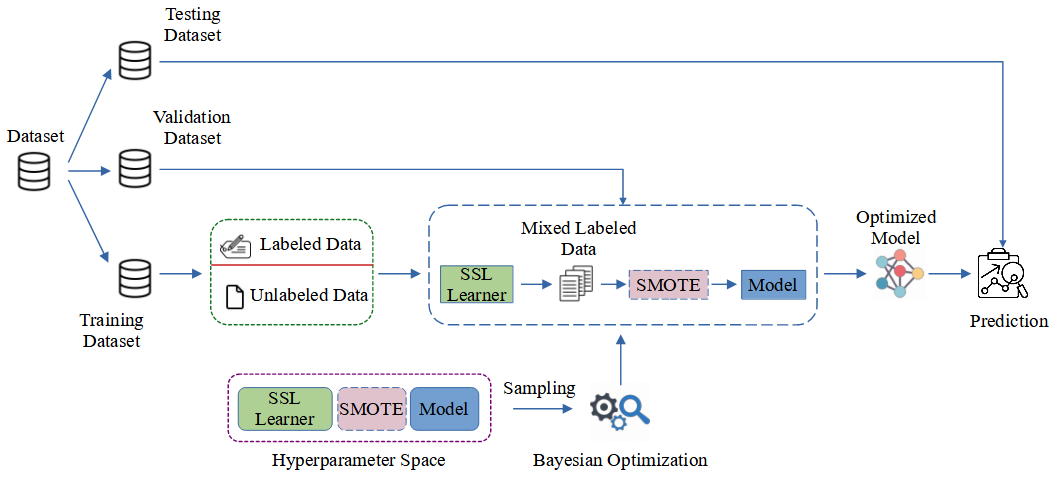}
\caption{An overview of the architecture of Dapper framework.}
\label{fig:dapper_architecture}
\end{figure*}


\subsection{Hyperparameter Optimization}\label{sec:hpo}
Hyperparameter is a type of parameter in machine learning models that can be estimated from data learning and have to be set before model training~\cite{yang2020hyperparameter}. Typically a hyperparameter has a known effect on a model in the general sense, but it is not clear how to best set a hyperparameter for a given dataset. In this sense, a range of possibilities have to be explored. \textit{Hyperparameter optimization} or \textit{hyperparameter tuning} is a technique that explores a range of hyperparameters and searches for the optimal solution for a task. 

Existing hyperparameter optimization methods can be mainly categorized into the following groups~\cite{yang2020hyperparameter}. The first group is \textit{decision-theoretic methods}. This kind of methods is based on the concept of defining a search space and select combinations in the search space. The most common methods of this type are grid search and random search. Grid search (GS)~\cite{bergstra2011algorithms} defines a search space as a grid of hyperparameter values and exhaustively searches and evaluates every position in the grid. Random search (RS)~\cite{bergstra2012random} defines a search space as a bounded domain of hyperparameter values and randomly samples points in that domain. Decision-theoretic methods are a good choice when the search space is small and not complicated.

Metaheuristic algorithms such as genetic algorithms and particle swarm optimization belongs to the second group. Genetic algorithms (GA)~\cite{lessmann2005optimizing} detect well-performing hyperparameter combinations during each generation, and pass them to the next generation until the optimal combination is found. In particle swarm optimization (PSO)~\cite{lorenzo2017particle}, each particle communicates with other particles to detect and update the current global optimum in each iteration until the final optimum is found. Metaheuristic algorithms suffer from time-consuming problems when search space is large since they are computationally expensive.

Unlike previous groups of methods, \textit{Bayesian optimization}~\cite{snoek2012practical,shahriari2015taking} is a novel hyperparameter optimization technique that keeps track of past evaluation results. The principle of Bayesian optimization is using those results to build a probability model of objective function, and maps hyperparameters to a probability of a score on the objective function, and therefore uses it to select the most promising hyperparameters to evaluate in the true objective function. This method is also called Sequential Model-Based Optimization (SMBO)~\cite{hutter2011sequential}. The probability representation of the objective function is called surrogate function or response surface because it is a high-dimensional mapping of hyperparameters to the probability of a score on the objective function. The surrogate function is much easier to optimize than the objective function and Bayesian methods work by finding the next set of hyperparameters to evaluate the actual objective function by selecting hyperparameters that perform best on the surrogate function. This method continually updates the surrogate probability model after each evaluation of the objective function.

\section{Methodology}~\label{sec:methodology}

Based on the above discussion, we were motivated to address our security problems using a combination of semi-supervised learning and hyperparameter optimization (and SMOTE when the data is imbalanced). The basis of our framework about semi-supervised learning is an idea called \textit{pseudo-labeling}~\cite{lee2013pseudo}. Pseudo-labeling works by iteratively propagating labels from labeled data to unlabeled data, i.e., relabeling the unlabeled data with algorithms. Our framework involves two pseudo-labeling approaches, i.e., \textit{label propagation} and \textit{label spreading}, which have been the \textit{de facto} standard method of the inference phase in graph-based semi-supervised learning. We first introduce both algorithms in detail and then present the proposed framework.

\subsection{Label Propagation}\label{sec:lpa}

Label propagation (LP) algorithm~\cite{zhu2002learning} is analogous to the $k$-Nearest-Neighbours algorithms and assumes that data points close to each other tend to have a similar label. To be specific, LP is an iterative algorithm that computes soft label assignments by pushing the estimated label at each node to its neighbouring nodes based on the edge weights. In other words, the new estimated label at each node is calculated as the weighted sum of the labels of its neighbours.

More formally, if we consider two set: $(x_1, y_1), ... (x_i, y_i) \in L$ as labeled datasets, and  $(x_{i+1}, y_{i+1}, ..., (x_n, y_n) \in U $ as unlabeled datasets, where $\{y_1, ..., y_n\} \in \{0, 1\}$ in a binary classification problem, and $\{x_1, ..., x_n\} \in \mathbb{R}$. LP constructs a graph $G = (V, E, W)$ where $V$ is the set of vertices representing set $L$ and $U$, and the edges in set $E$ represents the similarity of two node $i$ and $j$ with weight $W_{ij}$. The weight $W_{ij}$ is computed in a way that two nodes with smaller distance (i.e., more similar) will have a larger weight. Moreover, a Laplacian transition matrix on $V$ is denoted as 
\begin{equation}
    T_{ij} = \frac{W_{ij}}{\sum_{V_{k}\in N(V_{i})} W_{ik}}
\end{equation}
is used to propagate the labels.

There are two repeated steps involved in this algorithm until the label assignment process converges. LP starts with an initial assignment, which is random for the unlabeled data points and equal to the true labels for the labeled data points, then LP propagates labels from each node to the neighbouring nodes and then reset the predictions of the labeled data points to the corresponding true labels. LP finally converges to a harmonic function and this process can also be interpreted as a random walk with the transition matrix and stops when a labeled node is hit.

\subsection{Label Spreading}\label{sec:lsa}

Label spreading (LS) algorithm~\cite{zhou2003learning} is a variant to the label propagation algorithm. LS aims at minimizing a loss function that has regularization properties which is more robust to noise data. Instead of using Laplacian transition matrix for propagation, LS uses the normalized Laplacian matrix. A second different is the clamping effect LS has on the label distribution. Clamping allows LS to adjust the weight of the ground truth labeled data to some degree, rather than using hard clamping of input labels in LP.

\subsection{Dapper Framework}\label{sec:framework}

Figure~\ref{fig:dapper_architecture} presents the framework of Dapper. We split the original dataset int three parts, the training dataset, the validation dataset and the testing dataset with a predefined ratio $r_{1}$. The training dataset is further split into two subsets with another ratio $r_{2}$. One subset is denoted as the labeled training dataset $S_{L}$. We remove all the actual labels of the other subset $S_{U}$ and reset with default value $-1$. We treat subset $S_{U}$ as the unlabeled training dataset.

The optimization module in the Dapper framework has two sub-modules (as the blue dashed box shows in Figure~\ref{fig:dapper_architecture}), the semi-supervised learning algorithm and the machine learning classifier. Our framework is also \textit{adaptive}, which means when Dapper detects that the input training dataset is highly imbalanced (i.e., the percentage of minority class samples is lower than a predefined threshold $t$), Dapper automatically adds an oversampling sub-module. This oversampling sub-module is based on a tuned version of SMOTE which we discuss before in Section~\ref{sec:smote}.
The optimization process in the framework proceeds with a fixed number of evaluation trails, which is shown in Algorithm~\ref{alg:dapper}.

\begin{algorithm}
\small
    \SetKwInOut{Input}{Input}
    \SetKwInOut{Output}{Output}

    \underline{\textbf{Function} Dapper} $(D_{training}, D_{validation}, r, t, \theta, F)$\;
    \Input{Training dataset - $D_{training}$, \\
            Validation dataset - $D_{validation}$, \\
            SSL ratio - $r$, \\
            Imbalance threshold - $t$, \\
            Hyperparameter space - $\theta$, \\
            Target function - $F$}
    \Output{Optimized model $clf_{optimized}$}
    Split the training dataset into two subsets with ratio $r$, which treated as labeled dataset and unlabaled dataset \\ 
    Reset labels of unlabeled dataset to $-1$ \\
    \For{$trail_{i}$ $\in$ number of Bayesian Optimization trails}    
    { 
        Sample a combined hyperparameter set $\theta_{i}$ $\in$ $\theta$ \\ 
        Run SSL learner with $\theta_{i}$ and assign pseudo-labels \\
        Concatenate training subsets into $D_{mixed}$ with actual labels and pseudo-labels \\
        \If{ minority class percentage < $t$ }{
            Rebalance $D_{mixed}$ with SMOTE with $\theta_{i}$ \\
        }
        Train classifier with $D_{mixed}$ with $\theta_{i}$ \\ 
        Evaluate trained classifier with $D_{validation}$ \\
        Compute loss towards target function $F$ \\
    }
    Rank all optimization trails by loss with smallest on the top \\
    \Return Optimized model $clf_{optimized}$
    \caption{Pseudocode of Dapper's optimization process.}
    \label{alg:dapper}
\end{algorithm}

\begin{table*}[!t]
\centering
\small
\caption{Hyperparameter space explored in this study. This space covers two semi-supervised learning algorithms, SMOTE, the Random Forest classifier and the imbalance threshold.}
\begin{tabular}{c|c|c|l}
\hline
\textbf{Item} & \textbf{Hyperparameter} & \textbf{Range} & \multicolumn{1}{c}{\textbf{Brief Description}} \\ \hline \hline
\multirow{4}{*}{\begin{tabular}[c]{@{}c@{}}Label\\ Propagation\end{tabular}} & kernel & 'knn', 'rbf' & Kernel function to use. \\ 
 & gamma & (10, 30) & Parameter for rbf kernel. \\  
 & n\_neighbors & (5, 15) & Parameter for knn kernel. \\  
 & max\_iter & (500, 1500) & The maximum number of iterations allowed. \\ \hline
\multirow{5}{*}{\begin{tabular}[c]{@{}c@{}}Label\\ Spreading\end{tabular}} & kernel & 'knn', 'rbf' & Kernel function to use. \\  
 & gamma & (10, 30) & Parameter for rbf kernel. \\  
 & n\_neighbors & (5, 15) & Parameter for knn kernel. \\  
 & alpha & (0.1, 0.9) & Clamping factor. \\  
 & max\_iter & (500, 1500) & The maximum number of iterations allowed. \\ \hline
\multirow{3}{*}{SMOTE} & k & {[}1, 20{]} & Number of neighbours. \\  
 & r & {[}1, 6{]} & Minkowski distance metric. \\  
 & m & {[}50, 500{]} & Number of synthetic samples. \\ \hline
\multirow{7}{*}{\begin{tabular}[c]{@{}c@{}}Random\\ Forest\end{tabular}} & n\_estimators & {[}50, 200{]} & The number of trees in the forest. \\  
 & min\_samples\_leaf & {[}1, 25{]} & The minimum number of samples required to be at a leaf node. \\  
 & min\_samples\_split & {[}2, 25{]} & The minimum number of samples required to be at an internal node. \\  
 & max\_leaf\_nodes & {[}2, 100{]} & Total number of leaf nodes in a tree. \\  
 & max\_depth & {[}1, 25{]} & The maximum depth of the tree. \\  
 & max\_features & 'auto', 'sqrt', 'log2' & The number of features to consider when looking for the best split. \\  
 & bootstrap & 'True', 'False' & Whether bootstrap samples are used when building trees. \\ \hline
\begin{tabular}[c]{@{}c@{}}Imbalance\\ Threshold\end{tabular} & t & 30\% & A threshold to control whether SMOTE is used or not. \\ \hline
\end{tabular}
\label{tbl:sslRange}
\end{table*}

  During each trail, we repeat the following steps:
\begin{enumerate}
    \item With a chosen SSL learner (label propagation algorithm or label spreading algorithm), as well as a sampled combined set of hyperparameters, we assign pseudo-labels to unlabeled dataset with the algorithm.
    \item We concatenate the original labeled dataset and pseudo-labeled dataset into a new training dataset.
    \item If the percentage of the minority class samples of training dataset is lower than a threshold, we use SMOTE (with sampled hyperparameters) to balance the class ratio of the new training dataset. Otherwise, we pass this step.
    \item We train a classifier (with sampled hyperparameters) with the new training dataset, and evaluate the trained classifier on the validation dataset.
    \item The loss value of each trail, i.e. the complement of g-measure, is logged.
\end{enumerate}

After the optimization process, we rank all the evaluated classifiers by the loss value. The one with the smallest loss is selected and further tested on the testing dataset. Note that we sample the hyperparameters of SSL learner, SMOTE, and classifier in a combination manner, in which way we address multiple optimization problems simultaneously. Bayesian Optimization directs the whole sampling process and searches for the next promising set of hyperparameters after each trail from Table~\ref{tbl:sslRange}.

Besides, the reasons we design the Dapper framework in an adaptive manner to address the class imbalance issue are twofold. Firstly, data class imbalance is common in most security datasets, and many prior studies hint that oversampling the dataset is more likely to produce better classification performance. Secondly, we hypothesize that \textit{standard semi-supervised learning algorithms do not adequately address the class imbalance issue}. This means, in the mixed dataset with original labeled data and pseudo-labeled data, the problem still remains. Our experimental results further confirm the hypothesis, which we will discuss in detail in the result section.

In order to endorse the merits of Dapper (i.e., solving \textit{multiple optimization} problems), we also compare Dapper with two other treatments:
\begin{enumerate}
    \item \textit{No optimization:} with default SSL learner (default LP or default LS);
    \item \textit{Single optimization:} with optimized SSL learner only (optimized LP or optimized LS).
\end{enumerate}
The first treatment is used to endorse the merit of hyperparameter optimization, while the second treatment is used to demonstrate the advantages of Dapper over tuning SSL learners only. Moreover, our study performs sensitivity experiment, in which we pick different value of the ratio $r_2$ and explore the performance under each ratio. 



\section{Experiment}~\label{sec:Evaluation}

\subsection{Datasets and Algorithm}

Our proposed Dapper framework is evaluated with three security datasets which cover different security tasks, such as spam detection, malware detection and network intrusion detection.

\textbf{Twitter Spam}~\cite{chen20156}. As spam on Twitter becomes a growing problem, researchers have adopted different machine learning algorithms to detect Twitter spam. This dataset is generated from over 600 millions public tweets, and further labeled around 6.5 million spam tweets with 12 features extracted. The ground truth is established with Trend Micro's Web Reputation Service, which identify malicious tweets through URLs. We sample a total size of 5,000 instances from prior work~\cite{chen20156} (which has a size of about 100k), with 4,758 non-spam tweets and 242 spam tweets. This dataset has 12 features, such as account age, number of followers of the twitter user, the number of tweets the twitter user sent, etc.

\textbf{Malware URLs}~\cite{mamun2016detecting}. The original dataset collects about 114,400 URLs initially, containing benign and malicious URLs in four categories: spam URLs, phishing URLs, website URLs distributing malware and defacement URLs where pages belong to the trusted but compromised sites. This work selects malware URLs as our experimental target. In the selected dataset, more than 11,500 URLs related to malware websites were obtained from DNS-BH which is a project that maintain list of malware sites. There are 7,781 benign URLs and 6,711 malicious URLs in the dataset, and 79 features such as ratio of argument and URLs, count of token, and the proportion of digits in the URL parts, etc.

\textbf{CIC-IDS-2017}~\cite{sharafaldin2018toward}. This dataset consists of labeled network flows. It is comprised of both normal traffic and simulated abnormal data caused by intentional attacks on a test network. This dataset was constructed using the NetFlowMeter Network Traffic Flow analyzer, which collected multiple network traffic features and supported Bi-directional flows. We sample a portion of original dataset, which includes 11,425 normal traffic and 2,714 abnormal traffic. There are 70 features of the dataset, including average packet size, mean packet length, total forward packets, etc.

\begin{table}[!bp]
\centering
\small
\caption{Details of the studied dataset.}
\begin{tabular}{c|c|c|c|c}
\hline 
\textbf{Dataset} &
  \textbf{\begin{tabular}[c]{@{}c@{}}Training\\ Set\end{tabular}} &
  \textbf{\begin{tabular}[c]{@{}c@{}}Validation\\ Set\end{tabular}} &
  \textbf{\begin{tabular}[c]{@{}c@{}}Testing\\ Set\end{tabular}} &
  \textbf{\begin{tabular}[c]{@{}c@{}}Imbalance\\ Rate\end{tabular}} \\ \hline \hline
\begin{tabular}[c]{@{}c@{}}Twitter\\ Spam\end{tabular}   & 3,200 & 800 & 1,000 & 4.84\% \\ \hline
\begin{tabular}[c]{@{}c@{}}Malicious\\ URLs\end{tabular} & 9,274 & 2,319 & 2,899 & 46.3\% \\ \hline
CIC-IDS-2017 & 9,048 & 2,263 & 2,828 & 19.2\% \\ \hline
\end{tabular}
\label{tbl:dataset}
\end{table}

As we show in Table~\ref{tbl:dataset}, each dataset is split into training set, validation set and testing set with a ratio of 6.4 : 1.6 : 2 in a stratified way. To simulate the case of semi-supervised learning, we further divide the training set into labeled data and unlabeled data with different ratios in our sensitivity experiment. Another observation from Table~\ref{tbl:dataset} is the imbalance rate, where two datasets suffer from class imbalance issues. 

We select random forest classifier as our machine learning algorithm throughout the whole experiment. Random forest utilizes ensemble learning which consists of multiple decision trees. The `forest' generated by the algorithm is trained through bagging or bootstrap aggregating. Random forest establishes the outcome based on the predictions of the decision trees and predicts by taking the average or mean of the output from various decision trees. There are two reasons why we select random forest. Firstly, random forest is commonly used as classifier in previous security tasks such as intrusion detection~\cite{resende2018survey}. Secondly, the implementation of random forest in the Scikit-learn machine learning software library~\cite{scikit_learn} provides multiple hyperparameters that can be tuned (as can be seen from Table~\ref{tbl:sslRange}). 

Furthermore, the implementation of both label propagation algorithm and label spreading algorithm we adopt are publicly available in Scikit-learn, and the autotuned version of SMOTE is implemented according to ~\cite{agrawal2018better}.

\subsection{Evaluation Metrics}\label{sec:metrics}
If we let TP, TN, FP, FN to denote true positives, true negatives, false positives, and false negatives (respectively), we note that recall (pd), false positive rate (pf), g-measure (g-score), precision (prec), and f-measure (f1) are defined as follows:

\begin{equation}
    pd = \frac{TP}{TP + FN}
\end{equation}

\begin{equation}
    pf = \frac{FP}{FP + TN}
\end{equation}

\begin{equation}
    g-score = \frac{2 * pd * (100 - pf)}{ pd + (100 - pf)}
\end{equation}

\begin{equation}
    prec = \frac{TP}{TP + FP}
\end{equation}

\begin{equation}
    f1 = \frac{2 * pd * prec}{pd + prec}
\end{equation}
where 1) recall represents the ability of one algorithm to identify instances of positive class from the given dataset; 2) false positive rate measures the instances that are falsely classified by an algorithm as positive which are actually negative; 3) g-measure is the harmonic mean of recall and the complement of false positive rate, and it is also our optimization goal. We also report AUC-ROC value (i.e., Area Under the Receiver Operating Characteristics) for the completeness of results. This metric is an important metric to tell how much the model is capable to distinguish between classes. The higher, the better. In the worst case, the value is 0.5, which means the model has no discrimination ability between positive and negative class. Note that this study does not focus on other metrics such as precision or accuracy, as these metrics would fail to demonstrate the ability of a model under imbalanced classification. We report f-measure also for the completeness concern.

\section{Evaluation Results}~\label{sec:results}

\begin{table}[!bp]
\centering
\footnotesize
\caption{Summary results of 1) prior work which published the dataset; 2) 100\% training data used in a supervised way; and 3) 10\% of training dataset used with Dapper on label spreading. All results are based on the random forest classifier.}
\begin{subtable}{0.5\textwidth}
\sisetup{table-format=-1.2}   
\centering

\caption{Twitter Spam}\label{tab:sub_first}
\begin{tabular}{c|c|c|c}
\hline
\textbf{Metric} &
  \textbf{\begin{tabular}[c]{@{}c@{}}Results from\\ Prior Work~\cite{chen20156}\end{tabular}} &
  \textbf{\begin{tabular}[c]{@{}c@{}}100\% labeled\\ data used\end{tabular}} &
  \textbf{Dapper} \\ \hline \hline
Recall & 92.9 & 58.3 & 85.4 \\ 
\begin{tabular}[c]{@{}c@{}}False Positive\\ Rate\end{tabular} & 7.1 & 0.1  & 9.7  \\ 
G-measure & N/A & 73.6 & 87.8 \\ 
AUC-ROC & N/A & 79.1 & 87.8 \\ \
F-measure & 56.6 & 72.7 & 45.1 \\ \hline
\begin{tabular}[c]{@{}c@{}}Size of\\ Training Data\end{tabular} &
  \begin{tabular}[c]{@{}c@{}}A portion of\\ over 100k\end{tabular} &
  3200 &
  320 \\ \hline
\end{tabular}
\end{subtable}

\bigskip
\begin{subtable}{0.5\textwidth}
\sisetup{table-format=4.0} 
\centering
\caption{Malware URLs}\label{tab:sub_second}
\begin{tabular}{c|c|c|c}
\hline
\textbf{Metric} &
  \textbf{\begin{tabular}[c]{@{}c@{}}Results from\\ Prior Work~\cite{mamun2016detecting}\end{tabular}} &
  \textbf{\begin{tabular}[c]{@{}c@{}}100\% labeled\\ data used\end{tabular}} &
  \textbf{Dapper} \\ \hline \hline
Recall & 99.0 & 99.2 & 94.0 \\ 
\begin{tabular}[c]{@{}c@{}}False Positive\\ Rate\end{tabular} & N/A & 0.4  &  0.6 \\ 
G-measure & N/A & 99.4 & 96.6 \\ 
AUC-ROC & N/A & 99.4 & 96.7 \\ \
F-measure & 99.0 & 99.3 & 96.6 \\ \hline
\begin{tabular}[c]{@{}c@{}}Size of\\ Training Data\end{tabular} &
  9,274 & 9,274 & 927 \\ \hline
\end{tabular}
\end{subtable}

\bigskip
\begin{subtable}{0.5\textwidth}
\sisetup{table-format=1.2} 
\centering
\caption{CIC-IDS-2017}\label{tab:sub_third}
\begin{tabular}{c|c|c|c}
\hline
\textbf{Metric} &
  \textbf{\begin{tabular}[c]{@{}c@{}}Results from\\ Prior Work~\cite{sharafaldin2018toward}\end{tabular}} &
  \textbf{\begin{tabular}[c]{@{}c@{}}100\% labeled\\ data used\end{tabular}} &
  \textbf{Dapper} \\ \hline \hline
Recall & 97.0 & 98.3 & 96.3 \\ 
\begin{tabular}[c]{@{}c@{}}False Positive\\ Rate\end{tabular} & N/A  &  0.3 &  6.3 \\ 
G-measure & N/A & 99.0 & 95.0 \\ 
AUC-ROC & N/A & 99.0 & 95.0 \\ \
F-measure & 97.0 & 98.6 & 86.5 \\ \hline
\begin{tabular}[c]{@{}c@{}}Size of\\ Training Data\end{tabular} &
  \begin{tabular}[c]{@{}c@{}}A portion of\\ over 2.8m\end{tabular} &
  9,048 &
  904 \\ \hline
\end{tabular}
\end{subtable}
\label{tbl:summaryResult}
\end{table}

\begin{table}[!htbp]
\centering
\scriptsize
\caption{Results of the Twitter Spam dataset. The best results of each metric from different treatments in each label rate are highlighted in blue color. LP,LS= label propagation and label spreading (described in \S\ref{sec:lpa} and \S\ref{sec:lsa}).}
\begin{subtable}{0.5\textwidth}
\sisetup{table-format=-1.2}   
\centering

\caption{Recall}\label{tab:sub_first}
\begin{tabular}{l|ccccccccc}
\textit{\textbf{Label Rate}}   & 90\% & 80\% & 70\% & 60\% & 50\% & 40\% & 30\% & 20\% & 10\% \\ \hline
\textit{Default LP}   &  52.1    &   50.0   &    50.0  &   27.1   &   10.4   &   2.1   &  2.1    &   0.0   &   0.0   \\ 
\textit{Optimized LP} &   52.1   &   60.4   &  60.4   &  68.8    &  60.4    &  52.1    &   33.3   &   8.3   &   0.0   \\ 
\textit{\textbf{Dapper + LP}}  &  \cellcolor[HTML]{DAE8FC} 87.5    &   83.3   &   \cellcolor[HTML]{DAE8FC}85.4   &   81.3   &  85.4    &   83.3   &   83.3   &   83.3   &   72.9   \\ \hline
\textit{\textit{Default LS}}   &   58.3   &  60.4    &  56.3    &   68.8   &    66.7  &   68.8   &   50.0   &   43.8   &  29.2    \\ 
\textit{Optimized LS} &   58.3   &   62.5   &   58.3   &   68.8   &   68.8   &   68.8   &  66.7    &  52.1   &   39.6   \\ 
\textbf{\textit{Dapper + LS}}  &   \cellcolor[HTML]{DAE8FC}87.5   &   \cellcolor[HTML]{DAE8FC}87.5   &   \cellcolor[HTML]{DAE8FC}85.4   &   \cellcolor[HTML]{DAE8FC}87.5   &   \cellcolor[HTML]{DAE8FC}87.5   &   \cellcolor[HTML]{DAE8FC}85.4   &   \cellcolor[HTML]{DAE8FC}85.4   &   \cellcolor[HTML]{DAE8FC}85.4   &   \cellcolor[HTML]{DAE8FC}85.4   \\ 
\end{tabular}
\end{subtable}

\bigskip
\begin{subtable}{0.5\textwidth}
\sisetup{table-format=4.0} 
\centering
\caption{G-Measure}\label{tab:sub_second}
\begin{tabular}{l|ccccccccc}
\textbf{\textit{Label Rate}}   & 90\% & 80\% & 70\% & 60\% & 50\% & 40\% & 30\% & 20\% & 10\% \\ \hline
\textit{Default LP}   &  68.4    &   66.7   &   66.7   &   42.6   &  18.9    &   4.1   &   4.1   &   0.0   &   0.0   \\ 
\textit{Optimized LP} &   68.4   &   75.3   &   75.3   &  81.4    &    75.3  &   68.4   &   50.0   &   15.4   &   0.0   \\ 
\textbf{\textit{Dapper + LP}}  &   \cellcolor[HTML]{DAE8FC}89.1   &   87.3   &   88.7   &   86.7   &   89.2   &   \cellcolor[HTML]{DAE8FC}89.4   &  86.0    &   83.1   &   81.4   \\ \hline
\textit{Default LS}   &   73.7   &  75.3    &  72.0    &   81.4   &    79.9  &   81.4   &   66.6   &   60.8   &   45.2   \\ 
\textit{Optimized LS} &  73.7    &   76.9   &   73.7   &   81.4   &   81.4   &   81.4   &  79.9    &   68.4   &   56.7   \\ 
\textbf{\textit{Dapper + LS}}  &   \cellcolor[HTML]{DAE8FC}89.1   &   \cellcolor[HTML]{DAE8FC}89.1   &   \cellcolor[HTML]{DAE8FC}89.3   &  \cellcolor[HTML]{DAE8FC}88.3    &   \cellcolor[HTML]{DAE8FC}91.1   &   89.1   &   \cellcolor[HTML]{DAE8FC}88.9   &   \cellcolor[HTML]{DAE8FC}85.3   &   \cellcolor[HTML]{DAE8FC}87.8   \\ 
\end{tabular}
\end{subtable}

\bigskip
\begin{subtable}{0.5\textwidth}
\sisetup{table-format=1.2} 
\centering
\caption{AUC-ROC}\label{tab:sub_third}
\begin{tabular}{l|ccccccccc}
\textbf{\textit{Label Rate}}   & 90\% & 80\% & 70\% & 60\% & 50\% & 40\% & 30\% & 20\% & 10\% \\ \hline
\textit{Default LP}   &  76.0    &   75.0   &   75.0   &   63.5   &   55.2   &  51.0    &  51.0    &   50.0   &   50.0   \\ 
\textit{Optimized LP} &  76.0    &   80.2   &   80.1   &  84.3    &    80.1  &   75.9   &  66.7    &   54.2   &   50.0   \\ 
\textbf{\textit{Dapper + LP}}  &   \cellcolor[HTML]{DAE8FC}89.1   &  87.5    &   88.8   &    87.1  &   89.4   &   \cellcolor[HTML]{DAE8FC}89.9   &   86.0   &  83.1    &   82.5   \\ \hline
\textit{Default LS}   &   79.2   &   80.2   &   78.1   &   84.3   &    83.2  &  84.3    &  74.9    &   71.7   &  64.5    \\ 
\textit{Optimized LS} &  79.2    &   81.3   &   79.2   &   84.3   &  84.2    &   84.3   &   83.2   &   75.8   &   69.7   \\ 
\textbf{\textit{Dapper + LS}}  &  \cellcolor[HTML]{DAE8FC}89.1    &   \cellcolor[HTML]{DAE8FC}89.1   &  \cellcolor[HTML]{DAE8FC}89.5    &  \cellcolor[HTML]{DAE8FC}88.3    &  \cellcolor[HTML]{DAE8FC} 91.2   &  89.3    &   \cellcolor[HTML]{DAE8FC}89.1   &   \cellcolor[HTML]{DAE8FC}85.3   &   \cellcolor[HTML]{DAE8FC}87.8   \\ 
\end{tabular}
\end{subtable}

\bigskip
\begin{subtable}{0.5\textwidth}
\sisetup{table-format=4.0} 
\centering
\caption{False Positive Rate}\label{tab:sub_fourth}
\begin{tabular}{l|ccccccccc}
\textbf{\textit{Label Rate}}   & 90\% & 80\% & 70\% & 60\% & 50\% & 40\% & 30\% & 20\% & 10\% \\ \hline
\textit{Default LP}   &  \cellcolor[HTML]{DAE8FC} 0.0   &  \cellcolor[HTML]{DAE8FC}0.0    &   \cellcolor[HTML]{DAE8FC}0.0   &   \cellcolor[HTML]{DAE8FC}0.0   &  \cellcolor[HTML]{DAE8FC}0.0    &   \cellcolor[HTML]{DAE8FC}0.0   &   \cellcolor[HTML]{DAE8FC}0.0   &  \cellcolor[HTML]{DAE8FC}0.0    &   \cellcolor[HTML]{DAE8FC}0.0   \\ 
\textit{Optimized LP} &  \cellcolor[HTML]{DAE8FC}0.0    &   \cellcolor[HTML]{DAE8FC}0.0   &  0.2   &   0.2   &  0.2    &   0.2   &   \cellcolor[HTML]{DAE8FC}0.0   &   \cellcolor[HTML]{DAE8FC}0.0   &   \cellcolor[HTML]{DAE8FC}0.0   \\ 
\textbf{\textit{Dapper + LP}}  &  9.2    &   8.3   &   7.7   &   7.0   &   6.6   &  3.6   &   11.2   &  17.2    &   8.0   \\ \hline
\textit{Default LS}   &   \cellcolor[HTML]{DAE8FC}0.0   &  \cellcolor[HTML]{DAE8FC}0.0    &   0.1   &   0.2   &   0.3   &   0.2   &   0.1  &   0.3   &   0.1   \\ 
\textit{Optimized LS} &   \cellcolor[HTML]{DAE8FC}0.0   &   \cellcolor[HTML]{DAE8FC}0.0   &   0.0   &   0.2   &  0.3    &   0.3   &   0.3   &  0.5    &   0.2   \\ 
\textbf{\textit{Dapper + LS}}  &   9.2   &   9.2   &   6.5   &   10.9   &  5.0    &   6.8   &  7.2    &   14.8   &   9.7   \\ 
\end{tabular}
\end{subtable}

\bigskip
\begin{subtable}{0.5\textwidth}
\sisetup{table-format=4.0} 
\centering
\caption{F-Measure}\label{tab:sub_fifth}
\begin{tabular}{l|ccccccccc}
\textbf{\textit{Label Rate}}   & 90\% & 80\% & 70\% & 60\% & 50\% & 40\% & 30\% & 20\% & 10\% \\ \hline
\textit{Default LP}   &   68.5   &   66.7   &   66.7   &   42.6   &  18.9    &  4.1    &   4.1   &   0.0   &   0.0   \\ 
\textit{Optimized LP} &   68.5   &   75.3   &   73.4   &   \cellcolor[HTML]{DAE8FC}79.5   &    73.4  &   66.7   &   0.5   &  15.4    &  0.0   \\ 
\textbf{\textit{Dapper + LP}}  &  47.2    &   47.9   &   50.3   &  50.6    &  53.9    &   65.6   &   41.0   &   31.7   &   44.0   \\ \hline
\textit{Default LS}   &   \cellcolor[HTML]{DAE8FC}73.7   &   75.3   &   71.1   &   \cellcolor[HTML]{DAE8FC}79.5   &   77.1   &  \cellcolor[HTML]{DAE8FC}79.5    &   65.8   &  58.3    &   44.4   \\ 
\textit{Optimized LS} &  \cellcolor[HTML]{DAE8FC}73.7    &    \cellcolor[HTML]{DAE8FC}76.9  &  \cellcolor[HTML]{DAE8FC}73.7    &  \cellcolor[HTML]{DAE8FC}79.5    &  \cellcolor[HTML]{DAE8FC} 78.6   &   78.7   &   \cellcolor[HTML]{DAE8FC}77.1   &   \cellcolor[HTML]{DAE8FC}64.1   &   \cellcolor[HTML]{DAE8FC}55.1   \\ 
\textbf{\textit{Dapper + LS}}  &   47.2   &   47.2   &  54.3    & 43.3      &  60.9    &   53.2   &  51.9    &   35.7   &  45.1    \\ 
\end{tabular}
\end{subtable}
\label{tbl:spamResults}
\end{table}

\begin{table}[!htbp]
\centering
\scriptsize
\caption{Results of the Malware URLs dataset. The best results of each metric from different treatments in each label rate are highlighted in blue color. LP,LS= label propagation and label spreading (described in \S\ref{sec:lpa} and \S\ref{sec:lsa}).}
\begin{subtable}{0.5\textwidth}
\sisetup{table-format=-1.2}   
\centering

\caption{Recall}\label{tab:sub_first}
\begin{tabular}{l|ccccccccc}
\textbf{\textit{Label Rate}}   & 90\% & 80\% & 70\% & 60\% & 50\% & 40\% & 30\% & 20\% & 10\% \\ \hline
\textit{Default LP}   &  96.3    &   91.2   &   84.4   &  70.3    &   48.7   &  26.5    &   11.5   &   5.7   &   1.8   \\ 
\textit{Optimized LP} &  98.3    &  98.1    &  98.1    &  97.2  &   97.3   &  94.5     &   76.2   &  66.9    &   48.1   \\ 
\textbf{\textit{Dapper + LP}}  &   98.4   &  98.2    &   98.1   &   97.5   &   97.1   &  97.1  &   79.2   &   69.1   &   49.2   \\ \hline
\textit{Default LS}   &  98.2    &   98.3   &   98.3   &   98.1   &   \cellcolor[HTML]{DAE8FC}97.8   & 97.2     &  96.3    &  95.2    &   91.2   \\ 
\textit{Optimized LS} &  \cellcolor[HTML]{DAE8FC}99.0    &   98.6   &   98.0   &   97.9   &  97.6    &  97.5    &   \cellcolor[HTML]{DAE8FC}97.7   &   95.8   &   91.2   \\ 
\textbf{\textit{Dapper + LS}}  &   \cellcolor[HTML]{DAE8FC}99.0   &  \cellcolor[HTML]{DAE8FC} 98.8   &   \cellcolor[HTML]{DAE8FC}98.5   &   \cellcolor[HTML]{DAE8FC}98.4   &  \cellcolor[HTML]{DAE8FC}97.8    &   \cellcolor[HTML]{DAE8FC}98.0   &   97.1   &  \cellcolor[HTML]{DAE8FC}96.3    &   \cellcolor[HTML]{DAE8FC}94.0   \\ 
\end{tabular}
\end{subtable}

\bigskip
\begin{subtable}{0.5\textwidth}
\sisetup{table-format=4.0} 
\centering
\caption{G-Measure}\label{tab:sub_second}
\begin{tabular}{l|ccccccccc}
\textbf{\textit{Label Rate}}   & 90\% & 80\% & 70\% & 60\% & 50\% & 40\% & 30\% & 20\% & 10\% \\ \hline
\textit{Default LP}   &   97.9   &   95.3   &   91.5   &   82.6   &   65.5   &  41.8    &  20.6    &   10.7   &  3.5    \\ 
\textit{Optimized LP} &   98.1   &   98.7   &   98.4   &   97.7   &  97.9    &   96.6   &  86.4    &   80.1   &   64.9   \\ 
\textbf{\textit{Dapper + LP}}  &   98.8   &   98.7   &   98.5   &   98.3   &  98.0    &   97.7   &   88.2   &  81.7   &   65.9   \\ \hline
\textit{Default LS}   &   98.7   &   98.7   &   98.7   &   98.6   &   98.2   &  98.3    &   97.8   &   97.2   &   94.9   \\ 
\textit{Optimized LS} &   99.2   &  99.1    &   98.8   &   \cellcolor[HTML]{DAE8FC}98.7   &   98.4   &  98.3    &   97.9   &   97.4   &   94.9   \\ 
\textbf{\textit{Dapper + LS}}  &   \cellcolor[HTML]{DAE8FC}99.3   &   \cellcolor[HTML]{DAE8FC}99.2   &   \cellcolor[HTML]{DAE8FC}98.9   &   \cellcolor[HTML]{DAE8FC}98.7   &   \cellcolor[HTML]{DAE8FC}98.6   &   \cellcolor[HTML]{DAE8FC}98.4   &   \cellcolor[HTML]{DAE8FC}98.1   &  \cellcolor[HTML]{DAE8FC}97.7    &  \cellcolor[HTML]{DAE8FC} 96.6   \\ 
\end{tabular}
\end{subtable}

\bigskip
\begin{subtable}{0.5\textwidth}
\sisetup{table-format=1.2} 
\centering
\caption{AUC-ROC}\label{tab:sub_third}
\begin{tabular}{l|ccccccccc}
\textbf{\textit{Label Rate}}   & 90\% & 80\% & 70\% & 60\% & 50\% & 40\% & 30\% & 20\% & 10\% \\ \hline
\textit{Default LP}   &  97.9    &   95.5   &    92.1   &   85.1   &   74.4   &   63.2   &   55.7   &   52.8   &   50.9   \\ 
\textit{Optimized LP} &  98.1    &   98.7   &   98.4   &   97.7   &   97.9   &  96.7    &   88.0   &   83.4   &   74.0   \\ 
\textbf{\textit{Dapper + LP}}  &   98.8   &   98.7   &   98.5   &   98.3   &   98.0   &   97.8   &   89.4   &   84.5   &   74.5   \\ \hline
\textit{Default LS}   &   98.7   &   98.7   &  98.7    &  98.6    &  98.2    &  98.3    &  97.8    &   97.2   &   95.0   \\ 
\textit{Optimized LS} &   99.2   &  99.1    &  98.8    &   \cellcolor[HTML]{DAE8FC}98.7   &   98.4   &   98.3   &   97.9   &   97.5   &   95.0   \\ 
\textbf{\textit{Dapper + LS}}  &  \cellcolor[HTML]{DAE8FC}99.3    &  \cellcolor[HTML]{DAE8FC} 99.2   &  \cellcolor[HTML]{DAE8FC} 98.9   &  \cellcolor[HTML]{DAE8FC} 98.7   &   \cellcolor[HTML]{DAE8FC}98.6   &  \cellcolor[HTML]{DAE8FC}98.4    &   \cellcolor[HTML]{DAE8FC}98.1   & \cellcolor[HTML]{DAE8FC} 97.7    &   \cellcolor[HTML]{DAE8FC}96.7   \\ 
\end{tabular}
\end{subtable}

\bigskip
\begin{subtable}{0.5\textwidth}
\sisetup{table-format=4.0} 
\centering
\caption{False Positive Rate}\label{tab:sub_fourth}
\begin{tabular}{l|ccccccccc}
\textbf{\textit{Label Rate}}   & 90\% & 80\% & 70\% & 60\% & 50\% & 40\% & 30\% & 20\% & 10\% \\ \hline
\textit{Default LP}   &  \cellcolor[HTML]{DAE8FC}0.3    &   \cellcolor[HTML]{DAE8FC}0.2   &   \cellcolor[HTML]{DAE8FC}0.1   &   \cellcolor[HTML]{DAE8FC}0.0   &   \cellcolor[HTML]{DAE8FC}0.0   &   \cellcolor[HTML]{DAE8FC}0.0   &   \cellcolor[HTML]{DAE8FC}0.0   &   \cellcolor[HTML]{DAE8FC}0.1   &   \cellcolor[HTML]{DAE8FC}0.0   \\ 
\textit{Optimized LP} &   0.6   &   0.6   &   1.1   &  0.7   &   1.4   &     1.1 &  0.2   &   0.2   &   0.1   \\ 
\textbf{\textit{Dapper + LP}}  &  0.7    &   0.8   &  0.9    &   0.9   &  1.1    &   1.6   &   0.4   &  \cellcolor[HTML]{DAE8FC}0.1    &   0.1   \\ \hline
\textit{Default LS}   &   0.9   &  0.8    &   0.9   &   1.0   &  1.3    &   0.6   &   0.7   &  0.8    &   1.1   \\ 
\textit{Optimized LS} &  0.5    &  0.4    &   0.4   &  0.5    &   0.7   &   0.8   &  1.8    &   0.8   &   1.2   \\ 
\textbf{\textit{Dapper + LS}}  &  0.5    &  0.4    &  0.6    &   1.0   &  0.6    &   1.2   &   0.9   &   0.9   &   0.6   \\ 
\end{tabular}
\end{subtable}

\bigskip
\begin{subtable}{0.5\textwidth}
\sisetup{table-format=4.0} 
\centering
\caption{F-Measure}\label{tab:sub_fifth}
\begin{tabular}{l|ccccccccc}
\textbf{\textit{Label Rate}}   & 90\% & 80\% & 70\% & 60\% & 50\% & 40\% & 30\% & 20\% & 10\% \\ \hline
\textit{Default LP}   &  97.8    &   95.3   &    91.4   &   82.6   &   65.5   &  41.8    &   20.6   &   10.7   &   3.5   \\ 
\textit{Optimized LP} &  98.0    &   98.7   &   98.4   &   97.6   &   97.8   &   96.5   &   86.4   &   80.1   &   64.9   \\ 
\textbf{\textit{Dapper + LP}}  &  98.7    &  98.6    &   98.5   &   98.2   &  97.9    &   97.6   &   88.1   &   81.7   &   65.9   \\ \hline
\textit{Default LS}   &  98.6    &   98.7   &   98.6   &  98.5    &  98.1    &  98.2    &   97.7   &   97.1   &  94.7    \\ 
\textit{Optimized LS} &   \cellcolor[HTML]{DAE8FC}99.2   &  99.0    &   98.8   &  \cellcolor[HTML]{DAE8FC}98.6    &  98.3    &   98.2   &  97.8    &   97.3   &   94.7   \\ 
\textbf{\textit{Dapper + LS}}  &   \cellcolor[HTML]{DAE8FC}99.2   &  \cellcolor[HTML]{DAE8FC}99.2    &   \cellcolor[HTML]{DAE8FC}98.9   &   \cellcolor[HTML]{DAE8FC}98.6   &   \cellcolor[HTML]{DAE8FC}98.5   &   \cellcolor[HTML]{DAE8FC}98.3   &   \cellcolor[HTML]{DAE8FC}98.0   &   \cellcolor[HTML]{DAE8FC}97.6   &   \cellcolor[HTML]{DAE8FC}96.6   \\ 
\end{tabular}
\end{subtable}
\label{tbl:urlResults}
\end{table}

\begin{table}[!htbp]
\centering
\scriptsize
\caption{Results of the CIC-IDS-2017 dataset. The best results of each metric from different treatments in each label rate are highlighted in blue color. LP,LS= label propagation and label spreading (described in \S\ref{sec:lpa} and \S\ref{sec:lsa}).}
\begin{subtable}{0.5\textwidth}
\sisetup{table-format=-1.2}   
\centering

\caption{Recall}\label{tab:sub_first}
\begin{tabular}{l|ccccccccc}
\textbf{\textit{Label Rate}}   & 90\% & 80\% & 70\% & 60\% & 50\% & 40\% & 30\% & 20\% & 10\% \\ \hline
\textit{Default LP}   &  95.9    &   90.1   &   80.1   &  67.6  &    50.3  &  28.9    &   17.3   &   8.1   &  3.3    \\ 
\textit{Optimized LP} &   97.6   &  97.4    &  97.4    &   95.7   &  94.6    &   90.8   &   82.1   &   42.3   & 19.3\\ 
\textbf{\textit{Dapper + LP}}  &   \cellcolor[HTML]{DAE8FC}99.4   &   \cellcolor[HTML]{DAE8FC}99.8   &   \cellcolor[HTML]{DAE8FC}99.6   &   \cellcolor[HTML]{DAE8FC}99.1   &   98.9   &   97.6   &  97.6    &    96.8  &   95.0   \\ \hline
\textit{Default LS}   &  98.1    &   97.2   &   97.2   &  95.8    &    95.0  &   93.0   &   91.7   &   90.2   &   87.3   \\ 
\textit{Optimized LS} &  98.3    &   97.2   &   97.4   &  96.5    &   96.5   &   95.2   &  94.1    &   93.2   &   87.8   \\ 
\textbf{\textit{Dapper + LS}}  &   98.1   &   99.4   &   99.2   &  97.6    &   \cellcolor[HTML]{DAE8FC}99.4   &   \cellcolor[HTML]{DAE8FC}98.5   &   \cellcolor[HTML]{DAE8FC}98.0   &   \cellcolor[HTML]{DAE8FC}97.4   &   \cellcolor[HTML]{DAE8FC}96.3   \\ 
\end{tabular}
\end{subtable}

\bigskip
\begin{subtable}{0.5\textwidth}
\sisetup{table-format=4.0} 
\centering
\caption{G-Measure}\label{tab:sub_second}
\begin{tabular}{l|ccccccccc}
\textbf{\textit{Label Rate}}   & 90\% & 80\% & 70\% & 60\% & 50\% & 40\% & 30\% & 20\% & 10\% \\ \hline
\textit{Default LP}   &  97.9    &   94.7   &   88.9   &   80.7   &   66.9   &   44.8   &   29.5   &   15.0   &   6.4   \\ 
\textit{Optimized LP} &  98.7    &   \cellcolor[HTML]{DAE8FC}98.5   &  98.4    &  97.6    &   96.7   &   94.6   &   89.5   &   59.5   &   32.4   \\ 
\textbf{\textit{Dapper + LP}}  &   98.1   &   98.4   &   \cellcolor[HTML]{DAE8FC}98.5   &   \cellcolor[HTML]{DAE8FC}97.9   &  \cellcolor[HTML]{DAE8FC}97.9    &   \cellcolor[HTML]{DAE8FC}98.1   &  \cellcolor[HTML]{DAE8FC}98.3    &   \cellcolor[HTML]{DAE8FC}97.6   &   \cellcolor[HTML]{DAE8FC}96.3   \\ \hline
\textit{Default LS}   &   98.9   &   98.4   &   98.3   &   97.4   &   96.8   &   95.8   &   95.0   &   93.9   &   92.0   \\ 
\textit{Optimized LS} &   \cellcolor[HTML]{DAE8FC}99.1   &   98.4   &   98.4   &  \cellcolor[HTML]{DAE8FC}97.9    &  97.8    &   97.1   &   96.4   &  95.7    &   92.5   \\ 
\textbf{\textit{Dapper + LS}}  &  97.2    &   96.9   &  97.2    & 96.5     &   97.5   &   97.4   &  96.5    &   96.5   &   95.0   \\ 
\end{tabular}
\end{subtable}

\bigskip
\begin{subtable}{0.5\textwidth}
\sisetup{table-format=1.2} 
\centering
\caption{AUC-ROC}\label{tab:sub_third}
\begin{tabular}{l|ccccccccc}
\textbf{\textit{Label Rate}}   & 90\% & 80\% & 70\% & 60\% & 50\% & 40\% & 30\% & 20\% & 10\% \\ \hline
\textit{Default LP}   &  97.9    &   95.0   &   90.0   &  83.8    &  75.1    &  64.4    &   58.6   &   54.1   &   51.6   \\ 
\textit{Optimized LP} &   98.7   &   \cellcolor[HTML]{DAE8FC}98.5   &   98.4   &   97.6   &   96.7   &  94.8    &   90.3   &  71.0    &   59.6   \\ 
\textbf{\textit{Dapper + LP}}  &   98.1   &   98.4   &   \cellcolor[HTML]{DAE8FC}98.5   &   \cellcolor[HTML]{DAE8FC}97.9   &  \cellcolor[HTML]{DAE8FC}97.9    &   \cellcolor[HTML]{DAE8FC}98.1   &   \cellcolor[HTML]{DAE8FC}98.3   &   \cellcolor[HTML]{DAE8FC}97.6   &   \cellcolor[HTML]{DAE8FC}96.3   \\ \hline
\textit{Default LS}   &   98.9   &  98.4    &   98.3   &   97.5   &   96.8   &   95.8   &   95.1   &  94.1    &   92.3   \\ 
\textit{Optimized LS} &   \cellcolor[HTML]{DAE8FC}99.1   &   98.4   &   98.4   &   \cellcolor[HTML]{DAE8FC}97.9   &   97.8   &   97.1   &   96.4   &   95.7   &   92.8   \\ 
\textbf{\textit{Dapper + LS}}  &  97.2   &   97.0   &   97.3   &  96.6    &   97.5   &   97.5   &  96.5    &  96.5    &  95.0   \\ 
\end{tabular}
\end{subtable}

\bigskip
\begin{subtable}{0.5\textwidth}
\sisetup{table-format=4.0} 
\centering
\caption{False Positive Rate}\label{tab:sub_fourth}
\begin{tabular}{l|ccccccccc}
\textbf{\textit{Label Rate}}   & 90\% & 80\% & 70\% & 60\% & 50\% & 40\% & 30\% & 20\% & 10\% \\ \hline
\textit{Default LP}   &   \cellcolor[HTML]{DAE8FC}0.1   &   \cellcolor[HTML]{DAE8FC} 0.1     &  \cellcolor[HTML]{DAE8FC} 0.0   &  \cellcolor[HTML]{DAE8FC}0.0    &  \cellcolor[HTML]{DAE8FC}0.0    &   \cellcolor[HTML]{DAE8FC}0.0   &  \cellcolor[HTML]{DAE8FC}0.0    &  \cellcolor[HTML]{DAE8FC}0.0    &   \cellcolor[HTML]{DAE8FC}0.0   \\ 
\textit{Optimized LP} &   0.2   &  0.4    &   0.5   &   0.5   &  1.2    &   1.2   &   1.6   &   0.3   &   0.2   \\ 
\textbf{\textit{Dapper + LP}}  &   3.2   &  2.9    &   2.6   &   3.3   &  3.0    &   1.3   &   0.9   &   1.6   &    2.4  \\ \hline
\textit{Default LS}   &   0.2   &   0.4   &   0.6   &   0.8   &   1.3   &  1.3    &   1.5   &   2.1   &   2.7   \\ 
\textit{Optimized LS} &  0.2    &   0.4   &  0.6    &   0.5   &    0.8  &  1.0    &   1.2   &   1.7   &   2.2   \\ 
\textbf{\textit{Dapper + LS}}  & 3.7    &   5.4   &   4.7   &   4.5   &   4.4   & 3.6 &  4.8    &  4.5    &  6.3    \\ 
\end{tabular}
\end{subtable}

\bigskip
\begin{subtable}{0.5\textwidth}
\sisetup{table-format=4.0} 
\centering
\caption{F-Measure}\label{tab:sub_fifth}
\begin{tabular}{l|ccccccccc}
\textbf{\textit{Label Rate}}   & 90\% & 80\% & 70\% & 60\% & 50\% & 40\% & 30\% & 20\% & 10\% \\ \hline
\textit{Default LP}   &   97.7   &   94.5   &  88.9    &   80.7   &   66.9   &   44.8   &   29.5   &   15.0   &   6.4   \\ 
\textit{Optimized LP} &   98.2   &   \cellcolor[HTML]{DAE8FC}97.9   &  \cellcolor[HTML]{DAE8FC}97.6    &  96.7    &   94.7   &   92.7   &   87.0   &   59.1   &   32.2   \\ 
\textbf{\textit{Dapper + LP}}  &   93.4   &  94.1    &   94.7   &   93.0   & 93.5     &   \cellcolor[HTML]{DAE8FC}96.1   &  \cellcolor[HTML]{DAE8FC}96.8    &   \cellcolor[HTML]{DAE8FC}95.2   &   \cellcolor[HTML]{DAE8FC}92.6   \\ \hline
\textit{Default LS}   &  98.7   &   97.8   &    97.2    &   96.2   &   94.8   &   93.8   &  92.6    &   90.6   &   87.9   \\ 
\textit{Optimized LS} &   \cellcolor[HTML]{DAE8FC}98.8   &  97.8    &  97.4    &  \cellcolor[HTML]{DAE8FC}97.1    &   \cellcolor[HTML]{DAE8FC}96.5   &   95.5   &   94.5   &   93.0   &   89.2   \\ 
\textbf{\textit{Dapper + LS}}  &  91.9    &  89.5    &   90.7   &     90.2  &  91.3    &   92.2   &   89.7   &  90.1    &   86.5   \\ 
\end{tabular}
\end{subtable}
\label{tbl:idsResults}
\end{table}

Our study is structured around the following research questions:

\begin{RQ}
{\bf RQ1.} Can we use less labeled training data with default SSL?
\end{RQ}

This research question explores whether we can use less labeled training data than supervised learning with default semi-supervised learning algorithms. 
Specifically, we will try to build predictors using  $10, 20, 30...90, 100\%$ of the labelled data. 

To answer this question, we first present results from two baseline treatments: 1) results from prior works which published and evaluated the original datasets; 2) results from sampled datasets with 100\% labeled training datasets used in training. Table~\ref{tbl:summaryResult} presents the baseline results and all the results come from the same classifier, i.e., random forest. There are several notes about the results: 1) Some prior works did not report results of some metrics, e.g. g-measure, which are denoted as N/A in the table; 2) Some prior work did not present details of data split design, hence we have no idea the exact size of training dataset to get those results, hence we denote as a portion of the whole original data size; 3) All the results are from the same machine learning algorithm, i.e., random forest, and the summary results indicate that all the datasets are able to achieve good performance, e.g., about 90\% recall, except the sampled dataset from Twitter spam, which can only achieve about 60\% recall with 100\% labeled training data.

Table~\ref{tbl:spamResults}, Table~\ref{tbl:urlResults} and Table~\ref{tbl:idsResults} present the results of Twitter spam dataset, malware URLs dataset and CIC-IDS-2017 dataset, respectively. Each table reports the results from evaluation metrics as defined in Section~\ref{sec:metrics}. Note that we use different label rate of the training dataset, and we report the results from each label rate.

There are several observations from these results:
\begin{enumerate}
    \item For each datasets, with default label propagation (Default LP) algorithm, the recall results show a decreasing trend when the label rate decreases. When the label rate is as low as 10\%, the recall results of label propagation is close to zero. 
    \item The results of default label spreading (Default LS) vary among all datasets. For example, for Twitter spam in Table~\ref{tbl:spamResults}, label spreading algorithm can achieve as low as half of recall of results when compared with supervised learning with 100\% labeled training data in Table~\ref{tbl:summaryResult}. For the rest two datasets, the recall results are much better, as we can achieve about 90\% of original recall performance even the label rate is low as 10\%.
\end{enumerate}

\begin{Answer}{}{RQ1}
Default semi-supervised learning algorithm such as label propagation cannot achieve ideal performance when label rate is low, while label spreading algorithm is much better than label propagation, but still cannot compared with supervised learning results for some datasets.
\end{Answer}

\begin{RQ}
{\bf RQ2.} Will hyperparameter optimization on SSL help improve the results?
\end{RQ}

We also present the results of all datasets with optimized semi-supervised learning algorithms. In this treatment, we only apply Bayesian optimization to seach the hyperparameter space of label propagation and label spreading, and use default random forest algorithm as well as not using SMOTE.

The results of using optimized semi-supervised learning algorithms are also several folds:
\begin{enumerate}
    \item For the Twitter spam dataset, the optimized label propagation algorithm (Optimized LP) cannot improve the recall results when the label rate is as low as 10\%, while the optimized label spreading algorithm (Optimized LS) can improve the recall from 29.2\% to 39.6\%, but still cannot achieve 58.3\% recall with 100\% label dataset used, and even far below 92.9\% original recall from prior work~\cite{chen20156}.
    \item For the malware URLs dataset and CIC-IDS-2017 dataset, the improved performance over default label propagation (Default LP) is obvious, but still not enough. However, for the label spreading algorithm, the optimized improvement is slight, since the default label spreading algorithm (Default LS) already has achieved a high performance. 
    \item Furthermore, by comparing optimized label propagation and optimized label spreading from all result tables, it is not hard to say optimized label spreading algorithm outperforms optimized label propagation, especially when the label rate is low. For example, for the Twitter spam dataset, the recall results from optimized label spreading (Optimized LS) is 39.6\%, which is far better than 0.0\% recall from optimized label propagation (Optimized LP).
    \item For other metrics such as AUC-ROC and F-measure, the advantages of optimization is similar to the recall results over default SSL settings. Besides, optimized label spreading is also better than optimized label propagation in these metrics.
\end{enumerate}

\begin{Answer}{}{RQ2}
Compared with default settings, hyperparameter optimization on semi-supervised algorithms can alleviate the decreasing trend of metrics such as recall and g-measure in label propagation, but still not good enough. Besides, the optimized label spreading algorithm show slight improvement over default label spreading.
\end{Answer}

\begin{RQ}
{\bf RQ3.} Can we endorse the merits of the Dapper framework?
\end{RQ}

\begin{figure}[!tp]
\centering
\includegraphics[width=7cm]{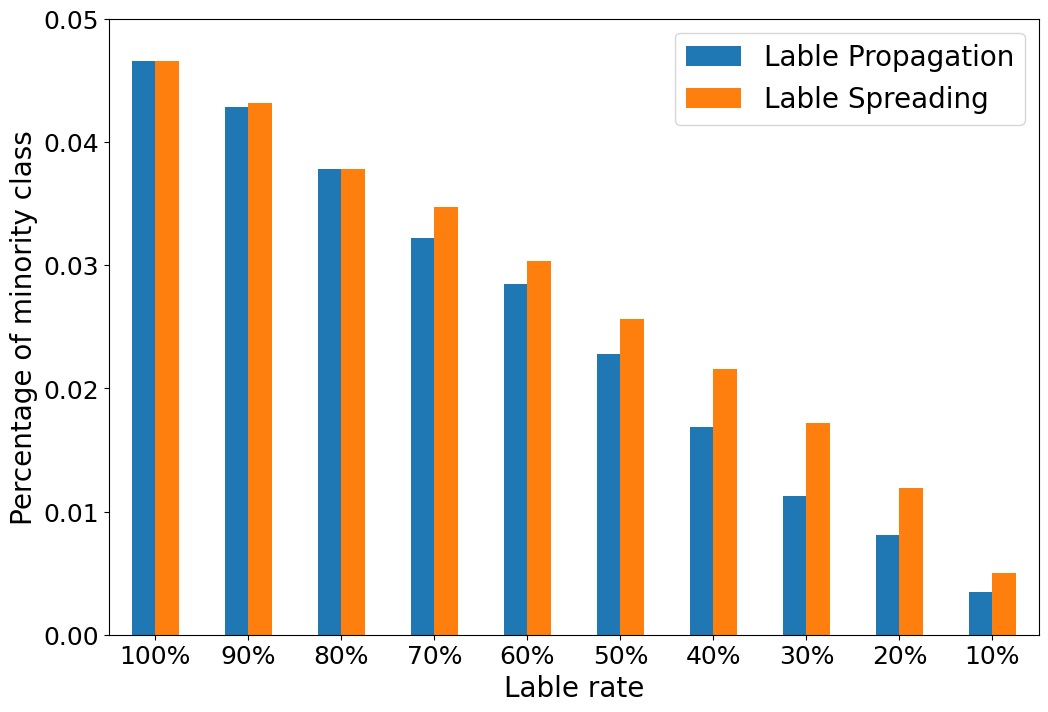}
\caption{The percentage of minority class under different label rates after using default label propagation and default label spreading in the Twitter spam dataset.}
\label{fig:imbalance_ratio}
\end{figure}

We now present the results of Dapper as we introduce before in Section~\ref{sec:framework}. Note that we hypothesize \textit{standard semi-supervised learning algorithms do not adequately address the class imbalance issue}. Figure~\ref{fig:imbalance_ratio} from the Twitter spam dataset validates our hypothesis. As we observe from the figure, the percentage of minority class in original labeled dataset is about 4.6\%. When decreasing the label rate, the imbalanced issue even get worse. For example, with only 10\% of labeled data, the ratio drops to about 0.5\% with label spreading and 0.3\% with label propagation. Several other prior studies also report similar findings~\cite{zhang2017label,iscen2019label}. This might result from the pseudo-labeling process, which infers the original minority class even to majority class. This finding also indicates the class imbalance issue still remains, and should not be ignored, which motivates us to add SMOTE to resample the mixed dataset in the adaptive Dapper framework.

Compared with default semi-supervised learning algorithms and optimized semi-supervised learning algorithms only, we make the following remarks on Dapper:

\begin{enumerate}
    \item In recall, g-measure and AUC-ROC, Dapper is more advantageous over other treatments in different levels. Even with label propagation, in the Twitter spam dataset and CIC-IDS-2017 dataset, Dapper can greatly increase the recall performance with 10\% label rate.
    \item As Dapper with label spreading (Dapper + LS) is slightly better than with label propagation (Dapper + LP) on Twitter spam and CIC-IDS-2017, but for malware URLs dataset, Dapper with label spreading is far better than with label propagation.
    \item What's more important, the results of Dapper is almost stable under different label rates.
    \item However, we have to note that, Dapper might bring in an increment of false positive rate (e.g., in Twitter spam and CIC-IDS-2017). Since the minority class ratio is lower than our pre-defined threshold, Dapper adaptively applies optimized SMOTE in the framework, which might cause the issue. We argue that, considering the improvement of important metrics such as recall, the trade-off of such increment in false positive rate is still acceptable.
\end{enumerate}

\begin{table}[!htbp]
\small
\centering
\caption{Average runtime (in minutes) of different treatments. We set the number of Bayesian Optimization trails to 100.}
\begin{tabular}{l|ccc}
\hline
\multirow{2}{*}{\textbf{Algorithm}} & \multicolumn{3}{c}{\textbf{Dataset}} \\ \cline{2-4} 
 & \multicolumn{1}{c|}{\textbf{\begin{tabular}[c]{@{}c@{}}Twitter\\ Spam\end{tabular}}} & \multicolumn{1}{c|}{\textbf{\begin{tabular}[c]{@{}c@{}}Malicious\\ URLs\end{tabular}}} & \textbf{CIC-IDS-2017} \\ \hline \hline
\textit{Default LP}                                                & \multicolumn{1}{c|}{< 1} & \multicolumn{1}{c|}{< 2} &  < 2 \\ 
\textit{Optimized LP}                                              & \multicolumn{1}{c|}{< 3} & \multicolumn{1}{c|}{< 10} & < 10 \\ 
\textbf{\textit{Dapper + LP}}                                               & \multicolumn{1}{c|}{< 4} & \multicolumn{1}{c|}{< 12} & < 15 \\ \hline
\textit{Default LS}                                                & \multicolumn{1}{c|}{< 1} & \multicolumn{1}{c|}{< 2} &  < 2 \\ 
\textit{Optimized LS}                                              & \multicolumn{1}{c|}{< 2} & \multicolumn{1}{c|}{< 6} &  < 6 \\ 
\textbf{\textit{Dapper + LS}}                                               & \multicolumn{1}{c|}{< 3} & \multicolumn{1}{c|}{< 10} & < 10 \\ \hline
\end{tabular}
\label{tbl:dapperRuntime}
\end{table}

Lastly, let's revisit the results from Table~\ref{tbl:summaryResult}. The last column of this table comes from Dapper, in which label spreading is selected as the SSL learner, and only 10\% of labeled training data is used. Compared with results from column 3 in which 100\% labeled data is used in a supervised paradigm, Dapper is close or even better in recall (with an acceptable trade-off in false positive rate). Compared with prior works which publish the dataset, Dapper shows obvious advantage in the labeled data size required. In addition, Table~\ref{tbl:dapperRuntime} shows the average runtime of different treatments, which also indicates the Dapper framework is also practical to use. The result also suggests that Dapper is a promising alternative to reduce the size of labeled data required to train a useful model.

\begin{Answer}{}{RQ3}
The adaptive Dapper framework with label spreading provides a close or even better performance than supervised learning with 100\% labeled training data, but with as low as 10\% of original labeled data required.
\end{Answer}

\section{Threats to Validity}~\label{sec:threats}


\textbf{Evaluation Bias.} In our work, we choose some popular evaluation metrics for classification tasks and use \textit{g-measure} as the optimization objective. We do not use other metrics because relevant information is not available to us, or we think they are not suitable enough for this specific task (e.g., precision).

\textbf{Optimizer Bias.} Dapper framework optimizes semi-supervised learning algorithm, machine learning classifier, or SMOTE with Bayesian Optimization. We do not claim Bayesian Optimization is the only best choice, but argue that Bayesian Optimization is fast to run and also more promising than other methods discussed in Section~\ref{sec:hpo}, and we believe Bayesian Optimization is good enough in our study.

\textbf{Learner Bias.} Research into automatic classifiers is a large and active field. Different machine learning algorithms have been developed to solve various classification problem tasks. Any data-mining study, such as this paper, can only use a small subset of the known classification algorithms. We select the random forest classifier, commonly used in similar classification tasks, for this work. In the future, we plan to explore more popular classifiers such as support vector machines (SVM), XGBoost~\cite{chen2015xgboost}, and so on.

\textbf{Implementation Bias.} The implementation of semi-supervised learning algorithms and the random forest classifier are from the Scikit-learn library, the implementation of Bayesian Optimization is from the hyperopt library~\cite{bergstra2013hyperopt}, and the implementation of the tunable SMOTE is from scratch by following the idea from ~\cite{agrawal2018better} without using existing available libraries. Different implementation of the above algorithms might have impact on the performance results, and might even change the conclusions from this work.

\textbf{Input Bias.} Our results come from the space of hyperparameter optimization listed in Table~\ref{tbl:sslRange}. In theory, other ranges might lead to other results. That said, our goal here is not to offer the {\em best} optimization but to argue that the optimized algorithms provided by Dapper can help reduce the ratio of labeled data required to a low degree, while still achieve promising performance, just by itself. For those purposes, we would argue that our current hyperparameter space suffices.

\section{Conclusion}~\label{sec:conclusion}

When labeled data is scarce, it can be hard to build adequately good prediction models. Prior works in software security have tried to address this issue with semi-supervised learning using a small pool of existing labels to infer the labels of unlabeled data. Those works usually do not explore SSL hyperparameter optimization (or even data rebalancing with SSL). This paper checks if that was a deficiency in prior works.

To perform that check, we propose Dapper that explores the hyperparameter space of existing semi-supervised learning algorithms, i.e., label propagation and label spreading, and machine learning classifier. When the percentage of minority class is low, Dapper further adaptively integrates an optimized oversampler SMOTE into the framework to address the class imbalance issue. Experimental results with three datasets show that Dapper's hyperparameter optimization and rebalancing combination can efficiently improve classification performance, even when most labels (90\%) are unavailable. In some datasets, we even observe better results with Dapper (using 10\% of the data) than using 100\% of all the labels. Based on those results, we recommend using hyperparameter optimization when dealing with label shortages for security tasks.  

\bibliographystyle{ACM-Reference-Format}
\bibliography{main}

\end{document}